\documentclass[aps,pre,twocolumn,groupedaddress]{revtex4-2}

\usepackage{amssymb}
\usepackage{amsmath}

\usepackage{dcolumn}%
\usepackage{bm}%
\usepackage{graphicx,textcomp}
\usepackage{array}
\usepackage{bm}
\usepackage{mathtools}
\usepackage{amssymb,amsfonts,amsmath}
\usepackage{color}
\usepackage{nicefrac}
\usepackage{algorithm}
\usepackage{algpseudocode}
\usepackage{enumerate}
\usepackage[all]{nowidow}
\usepackage{booktabs}
\usepackage{multirow}
\usepackage{xcolor}
\usepackage{lipsum}
\usepackage{hyphenat}
\begin{document}

\title{Model-based reconstruction of real-world fractal complex networks}

\author{Kordian Makulski, Mateusz J. Samsel, Micha\l~\L epek, Agata Fronczak, Piotr Fronczak} 

\affiliation{Faculty of Physics, Warsaw University of Technology, Koszykowa 75, Warsaw, PL-00-662, Poland}
\date{\today}
\begin{abstract}
This paper presents a versatile model for generating fractal complex networks that closely mirror the properties of real-world systems. By combining features of reverse renormalization and evolving network models, the proposed approach introduces several tunable parameters, offering exceptional flexibility in capturing the diverse topologies and scaling behaviors found in both natural and man-made networks. The model effectively replicates their key characteristics such as fractal dimensions, power-law degree distributions, and scale-invariant properties of hierarchically nested boxes. Unlike traditional deterministic models, it incorporates stochasticity into the network growth process, overcoming limitations like discontinuities in degree distributions and rigid size constraints. The model's applicability is demonstrated through its ability to reproduce the structural features of real-world fractal networks, including the Internet, the World Wide Web, and co-authorship networks.
\end{abstract}

\maketitle

\section{Introduction}

\subsection{State of the art}\label{SecIntroA}

Fractal complex networks constitute a remarkable and intriguing class of complex networks that are distinguished by their self-similarity and power-law scaling properties across a wide range of scales \cite{Song_2005, Song_2006, Fronczak_2024}. These networks are not confined to theoretical constructs; they appear ubiquitously in both natural and man-made systems. Examples include biological networks, such as neuronal or vascular systems, where structural and functional organization often reflects fractal-like patterns \cite{Gallos_2012}, the World Wide Web, which demonstrates hierarchical and scale-free properties \cite{Song_2005}, and various social networks~\cite{Gallos_2013,Fronczak_2022}, where communities often form self-similar clusters.

Modeling fractal networks is crucial to discovering the fundamental principles that govern the organization and evolution of these systems. Fractal network models provide a framework for studying processes such as information flow \cite{Gallos_2007}, robustness \cite{Chen_2019}, and spreading dynamics \cite{Goh_2006, Deppman_2021}, which are deeply influenced by the network's structure. By accurately modeling fractal networks, we can design more efficient and resilient infrastructures, predict behaviors in natural and engineered systems, and better understand the interplay between topology and function in such environments.

The primary feature that distinguishes fractal networks from non-fractal ones is that, when their nodes are grouped into non-overlapping boxes such that the maximum distance between any two nodes within the same box does not exceed $l_B$, the optimal number of such boxes exhibits power-law scaling \cite{Song_2005}: 
\begin{equation}\label{NblB} 
	N_B(l_B)\simeq N\,l_B^{-d_B}, 
\end{equation}
where $N$ is the number of nodes in the network and $d_B$ is its fractal (or box-counting) dimension. A direct consequence of this scaling (for $l_B = L$, when $N_B(L)=1$) is the power-law relationship between the network’s diameter $L$ and its size $N$, expressed as $L \sim N^{1/d_B}$ \cite{Rozenfeld_2010, Fronczak_2024}. In contrast, non-fractal networks typically exhibit a logarithmic relationship, $L \sim \log N$ \cite{SW1, SW2, SW3}, and are commonly referred to as small-world networks due to the extremely short average distances between their nodes.

When defining complex fractal networks, one must proceed with care, as there exist fractal networks that are not complex—for example, road networks \cite{road_2021} or random graph models at the percolation threshold \cite{Song_2005}. Conversely, there are also complex networks that exhibit scale invariance in their degree distributions—such as the well-known Barab\'asi-Albert (BA) model \cite{BAnet}—which, however, due to their small-world nature, are not structurally self-similar under renormalization~\cite{Song_2006}.

\begin{figure*}[t]
	\centering
	\includegraphics[width=0.6\textwidth]{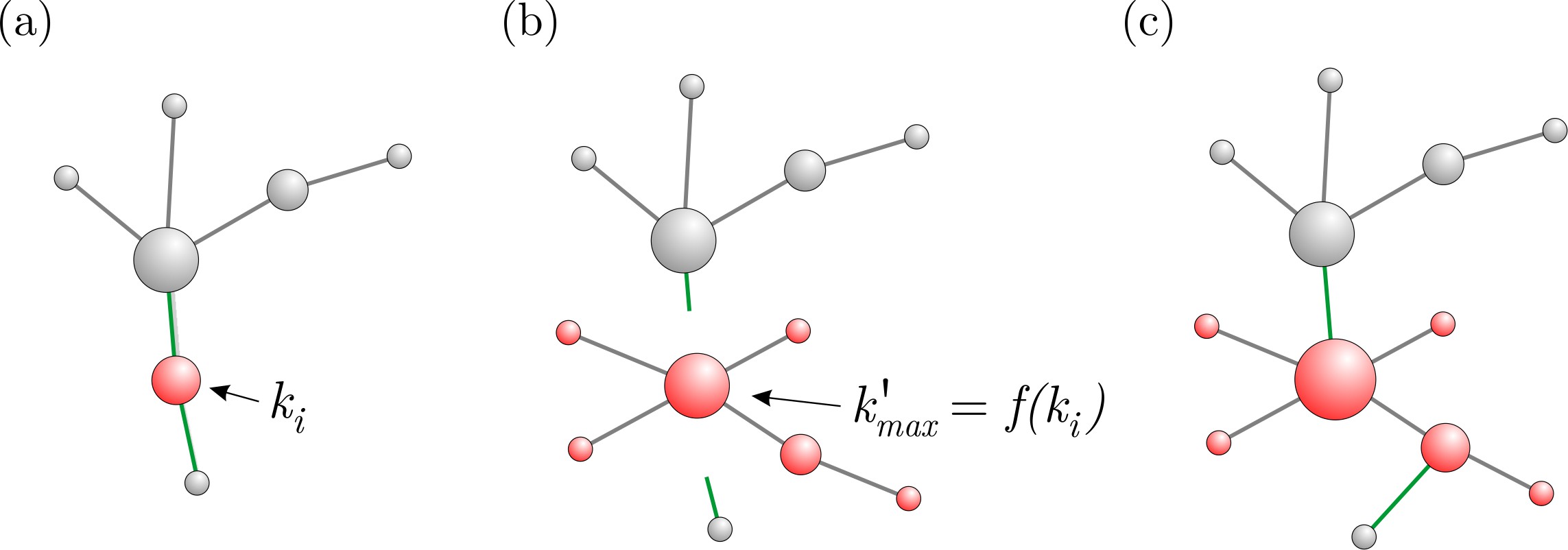}
	\caption{Illustration of the asynchronous replacement of nodes with subnetworks and their nesting within the target network to construct fractal structures (shown here for fractal trees with $a=1$): (a) a node $i$ with degree $k_i=2$ is randomly selected; (b) a new subnetwork with maximum degree $k'_{max}=f(2)=4$ is embedded at the position of node $i$; (c) external links are reconnected preferentially.} \label{Fig_schema}
\end{figure*}

The aforementioned renormalization is a process in which the non-overlaping boxes of a predefined diameter $l_B$ covering the fractal network are treated as super-nodes and used to create a new network \cite{Song_2005}. Remarkably, fractal networks preserve their statistical properties through successive renormalizations, exhibiting self-similarity that can be quantitatively analyzed \cite{Rozenfeld_2010, Fronczak_2024, Radicchi_2008}. For example, both the node degree distribution
\begin{equation}\label{Pk}
	P(k)\sim k^{-\gamma}, 
\end{equation}
and the box mass distribution
\begin{equation}\label{Pm} 
	P(m)\sim m^{-\delta}, 
\end{equation}
retain their power-law character before and after renormalization.

In \cite{Fronczak_2024}, it was shown that this structural stability under renormalization stems from the scale-invariant properties of the boxes themselves, whose masses depend not only on their diameter $l_B$, but also on the degree $k_B$ of the most connected node (the local hub) within each box:
\begin{equation}\label{mlBkB}
	m(l_B,k_B)\sim l_B^\alpha k_B^\beta,
\end{equation}
where $\alpha$ and $\beta$ are the so-called microscopic scaling exponents that characterize the local structure of fractal complex networks. Remarkably, the microscopic ($\alpha$ and $\beta$) and macroscopic ($d_B$, $\gamma$, and $\delta$) scaling exponents are directly related to each other through the following relations:
\begin{equation}\label{ab}
	\alpha=\frac{\delta-2}{\delta-1}d_B,\;\;\;\;\;\;\;\beta=\frac{\gamma-1}{\delta-1},
\end{equation}
which can be used to study how the local, microscopic structure influences global properties of these networks. In particular, as shown in \cite{Fronczak_2024}, the box-counting dimension is given by:
\begin{equation}\label{dB0}
	d_B=\alpha+\beta d_k,
\end{equation}
where $d_k$ is yet another microscopic exponent that describes how node degrees transform under renormalization:
\begin{equation}\label{dk}
	q=l_B^{-d_k}k_{B},
\end{equation}
where $q$ is the degree of the node which, during network renormalization using boxes with diameter $l_B$, replaces the box whose best-connected node had degree $k_B$, cf. Eq.~(\ref{mlBkB}). This exponent was originally introduced in the context of the scaling relation \cite{Song_2005, Yook_2005}:
\begin{equation}\label{gamma0}
	\gamma=1+\frac{d_B}{d_k},
\end{equation}
which established a direct link between the macroscopic properties of the network—its degree distribution and fractal dimension—and its underlying microscopic structure, thereby laying the groundwork for a unified scaling theory of fractal complex networks. These features distinguish such networks from other types, including small-world networks, which lack a scale-invariant local structure and therefore do not exhibit self-similarity under renormalization.

To avoid misunderstandings, in this work, we define complex fractal networks as those that: i. possess a well-defined fractal dimension $d_B$ (\ref{NblB}), ii. reveal scale-free property in terms of power-law distributions $P(k)$~(\ref{Pk}) and $P(m)$~(\ref{Pm}), and iii. are structurally self-similar (\ref{mlBkB}) under renormalization.  

\subsection{Modeling challenges}\label{SecIntroB}

Although the first models of fractal networks appeared at the beginning of this century, they remain relatively scarce and limited in their range of properties~\cite{Rosenberg_2020_book}. Despite the inherent randomness in the evolution of real fractal networks, most existing models are deterministic \cite{Song_2006, Zeng_2017_sierpinski, Zeng_2021_sturmian, Le_2015_sierpinski, Huang_2023_pentagon}. In a few cases—such as models based on random walks \cite{Chelminiak_2013, Ikeda_2020}—the construction method appears somewhat artificial and difficult to justify in light of the underlying processes occurring in real-world networks.

Deterministic models typically rely on reverse renormalization, wherein a single node (or edge) representing a unit at a higher scale is replaced by a more complex structure that better reflects the detailed properties of the original node or edge. Models in this class, such as the Song-Havlin-Makse (SHM) model \cite{Song_2006} and the $(u,v)$-flower model \cite{Rozenfeld_2007}, are analytically tractable and allow for precise verification of Eqs.~(\ref{NblB})-(\ref{ab}) \cite{Fronczak_2024}. However, they also introduce artificial discontinuities—such as abrupt jumps in the degree distribution and strict limitations on network size—which cannot be adjusted freely. For instance, the SHM model generates scale-free fractals with a tree-like structure, where the fractal dimension $d_B$ and the degree distribution exponent $\gamma$ are interdependent. In contrast, the $(u,v)$-flower model replaces edges with fixed-path subnetworks controlled by the parameters $u$ and $v$, thereby enabling a range of fractal configurations.

While these deterministic models are mathematically elegant, their limitations constrain their applicability to real-world systems, which often exhibit randomness and lack strict structural constraints. Recently, two models have been proposed that generalize the SHM and $(u,v)$-flower frameworks in an attempt to overcome these limitations \cite{Zakar_2022, Yakubo_2022}. The first model \cite{Zakar_2022} introduces randomness into the SHM framework, preserving the fractal nature of the network in a way that earlier attempts could not. The second model \cite{Yakubo_2022} constructs the network iteratively by replacing each edge of the previous generation with a smaller graph, known as a generator. By selecting different generators, this approach allows independent control over the scale-free property, fractality, and other structural characteristics of the network. Furthermore, the introduction of stochasticity renders this model a promising candidate for the further development of fractal network models. However, it still replicates certain issues inherent in reverse renormalization-based models, such as discontinuities in the distributions $P(k)$ and $P(m)$, as well as strictly defined network sizes resulting from a fixed number of renormalization steps.

In this paper, we propose a new model of fractal networks that lies at the interface between reverse renormalization-based models and evolving network models. Unlike deterministic alternatives, our approach introduces randomness at every stage of network construction, eliminating artificial discontinuities. The proposed model allows for flexible adjustment of key network parameters, including fractal dimension $d_B$ and the characteristic exponent of the degree distribution $\gamma$. These properties make the model particularly suitable for constructing synthetic networks that closely resemble different real-world fractal networks. 

The remainder of this paper is organized as follows: Section~\ref{SecModel} introduces the construction procedure of the model. Section~\ref{SecDatasets} presents real-world datasets used for comparisons. Section~\ref{SecResults} discusses the main results, highlighting the model's ability to replicate key features of real-world networks. Finally, Section~\ref{SecSummary} concludes the study.

\section{Model of an evolving fractal network}\label{SecModel}

The network construction procedure consists of the following stages:
	\begin{enumerate}
		\item[1.] Network growth begins from an initial configuration of nodes and links, which is not essential for the final structure of the network when it becomes sufficiently large. We assume that the initial configuration consists of two nodes connected by an edge.
		\item[2.] At each time step, a node $i$ with degree $k_i$ is randomly selected from the existing network and replaced by a subnetwork, according to the following procedure (cf. Fig.~\ref{Fig_schema}):
		\begin{enumerate}
			\item[a.] First, a subnetwork whose size depends on $k_i$ is constructed using the well-known generalized linear preferential attachment rule (g-PAR) \cite{Dorogovtsev_2000}. (This stage is described in more detail below.)
			\item[b.] Once the subnetwork has been fully generated, the $k_i$ edges that originally connected node $i$ to the rest of the network are reassigned—using the same g-PAR rule—to nodes within the newly created subnetwork.
		\end{enumerate}		
		\item[3.] Network growth terminates when the total number of nodes reaches the target size $N$.
	\end{enumerate}

With respect to stage 2a, the construction of the subnetwork that replaces node $i$ proceeds as follows:
\begin{enumerate}
	\item[i.] The subnetwork is initiated by a single node, to which we assign the so-called initial attractiveness \( A \in \mathbb{R}_+ \).
	\item[ii.] New nodes are added one by one, each making \( a \geq\!1 \) connection attempts to nodes already present in the subnetwork. An attempt may fail if it targets a node already connected, as no retries are made. Consequently, the actual number of created edges may be smaller than \( a \), especially in early stages when few targets are available.
	\item[iii.] Target nodes for connection attempts are selected according to the generalized preferential attachment (g-PAR) rule: the probability of choosing node \( j \) is proportional to \( k'_j + A \), where \( k'_j \) denotes the internal degree of node \( j \) within the subnetwork~\cite{Dorogovtsev_2000}.
	\item[iv.] The growth process continues until the most connected node in the subnetwork (i.e., the local hub) reaches a maximum internal degree \( k'_{max} = f(k_i) \), or when the number of nodes in the subnetwork exceeds the threshold \( n_{max} \).
\end{enumerate}

Note that the above construction procedure, in which the parameter $a$ is defined as a positive integer, can be extended by treating $a$ as the average number of edges. We comment on this in the Supplementary Material (SM).

Below, we briefly discuss the main assumptions underlying the model, with a focus on their individual roles and the rationale for introducing each of them.

Firstly, although the model inherits the concept of reverse renormalization, it proceeds via local, one‐node replacements at each time step rather than a global, synchronous update. This asynchronous scheme naturally gives rise to two distinct time scales: a global scale during which source nodes are chosen, and a local scale over which each selected node is expanded into its corresponding subnetwork.

Central to our inverse‐renormalization mechanism is the mapping
\begin{equation}\label{fk}
k'_{max} = f(k_i) = k_i^{\tau},
\end{equation}
which sets the maximum internal degree of the hub within the subnetwork replacing a node of degree \(k_i\). For \(\tau>1\), iterative application of \(f\) across successive reverse‐renormalization steps yields a nested, multi‐level hierarchy that faithfully reproduces the scale‐free and scale‐invariant properties observed in empirical fractal networks. However, this mapping has an important limitation: as the process continues and ever higher‐degree nodes emerge, the subnetworks that replace them grow correspondingly larger. Such subnetworks—if not subsequently replaced by further subnetworks—exhibit small‐world characteristics rather than true fractality, and their potentially large size can undermine the network’s global self‐similarity. To prevent this, we introduce an additional structural cutoff parameter, \(n_{max}\), which caps the maximum allowable size of each subnetwork.

Finally, control over the degree‐distribution exponent \(\gamma\) is provided by the initial attractiveness parameter \(A\). Drawing on the generalized preferential‐attachment rule of Dorogovtsev et al.~\cite{Dorogovtsev_2000}, varying \(A\) adjusts the relative likelihood that new nodes attach to existing hubs, thereby shaping the heavy tail of \(P(k)\).

In summary, the model is governed by five parameters: \(N\), \(a\), \(A\), \(\tau\), and \(n_{max}\). \(N\) sets the final network size, while \(a\) controls network density by specifying the average degree as \(\langle k\rangle = 2a\) (each new node introduces \(a\) edges, yielding an average degree of \(2a\)). The parameter \(A\) shapes the heavy tail of the degree distribution, tuning the exponent \(\gamma\) via the generalized preferential‐attachment mechanism. Finally, \(\tau\) and \(n_{max}\) determine the fractal dimension \(d_B\) by governing the form and size cutoff of the mapping \(f(k_i)\).

In Sec.~\ref{SecResultsA}, we investigate how independently varying \(A\), \(a\), and \(n_{max}\) affects both the macroscopic and microscopic properties of the model. Throughout most of this study, we keep \(\tau = 2\) fixed to ensure consistent nesting depth across simulations. Detailed explanation of this choice is provided in SM (see Fig.~S1). The role of this parameter is also revisited in Sec.~\ref{SecResultsB}, where \(\tau\) is released as a free parameter to fine‐tune the fit to empirical fractal networks (see Table~\ref{tablenetworks} for the adjusted values).

\begin{table}[t]
	\begin{center}
		\begin{tabular}{ | c | c | c | c | c | c | c | c | c | c | }
			\hline
			Network  & $N$     & $\langle k \rangle $ & $d_B$ & $\gamma$ & $\delta$& $A$ & $a$ & $\tau$ & $n_{max}$ \\ \hline \hline
			AS-Caida & 77,339  & 12.7                 & 5.2   & 2.1      &  2.2    &0.95& 6.35& 1.5      &40,000\\	\hline
			WWW& 855,802 & 10.0                 & 3.7   & 2.5      &  2.6    &2.5 & 5   & 2.0      &1,000\\ \hline
			DBLP     & 4,518   & 2.6                  & 2.5   & 2.6      &  3.3    & 0.65 & 1.3 & 1.3      &40 \\ \hline
		\end{tabular}
		\caption{Values of the parameters of the real-world networks used in this study. In the table, $N$ denotes the number of nodes in each network, $\langle k \rangle$ is the average node degree, and $d_B$, $\gamma$, and $\delta$ are macroscopic scaling exponents characterizing fractal complex networks. The remaining quantities, $A$, $a$, $\tau$, and $n_{max}$, are parameters of the proposed model.}\label{tablenetworks}
	\end{center}
\end{table}

\section{Datasets} \label{SecDatasets}

To demonstrate the usefulness of the proposed model in reflecting real-world networks, we analyzed the key characteristics of three different networks that have disctinct fractal exponents:

\begin{itemize}
	
	\item Internet at the level of autonomous systems: In context of the Internet, an autonomous system (AS) is a collection of associated Internet Protocol (IP) prefixes with a clearly defined routing policy. It governs how the AS exchanges routing information with other autonomous systems. A single AS can be thought of as a connected group of IP networks which are managed by a single administrative entity, e.g. a university, government, commercial organization or other type of internet service provider. The AS network ('AS-Caida') analyzed in this study contains 77.3 k nodes, a graph derived by CAIDA \cite{caida_as} from a set of RouteViews \cite{routeviews} BGP table snapshots from 1 November 2024. We obtained this network from public repository at \cite{stanford_caida_repo}. 
	
	\item WWW (Google web graph): The web subset analysed consists of 856 k web pages that are linked if there is a URL link from one page to another \cite{Leskovec_2009}. The dataset is publicly available in several network repositories (e.g. \cite{Rossi_2015_repo}).
	
	\item DBLP coauthorship network: DBLP is a digital library of article records published in computer science \cite{Tang_2012, DBLP_repo}. In this study, similarly as in Refs.~\cite{Fronczak_2022,Fronczak_2024}, we use the 12th version of the dataset (DBLP-Citation-network V12; released April 2020, which contains information on approximately 4.9 M articles published mostly during the last 20 years). We ourselves processed the raw DBLP data into the form of coauthorship network, from which we extracted the network backbone by imposing a threshold on the minimum number of joint papers ($\ge22$) two scientists should have. This procedure significantly reduced the size of the studied network (from 2.9 M nodes and 12.5 M links to 4.5 k nodes and 5.8 k edges), but thanks to it the network became naturally fractal.
\end{itemize}

The main characteristics of these networks, along with the model parameters used to generate their synthetic counterparts, are summarized in Table~\ref{tablenetworks}.

\begin{figure}[t]
	\centering
	\includegraphics[width=0.95\columnwidth]{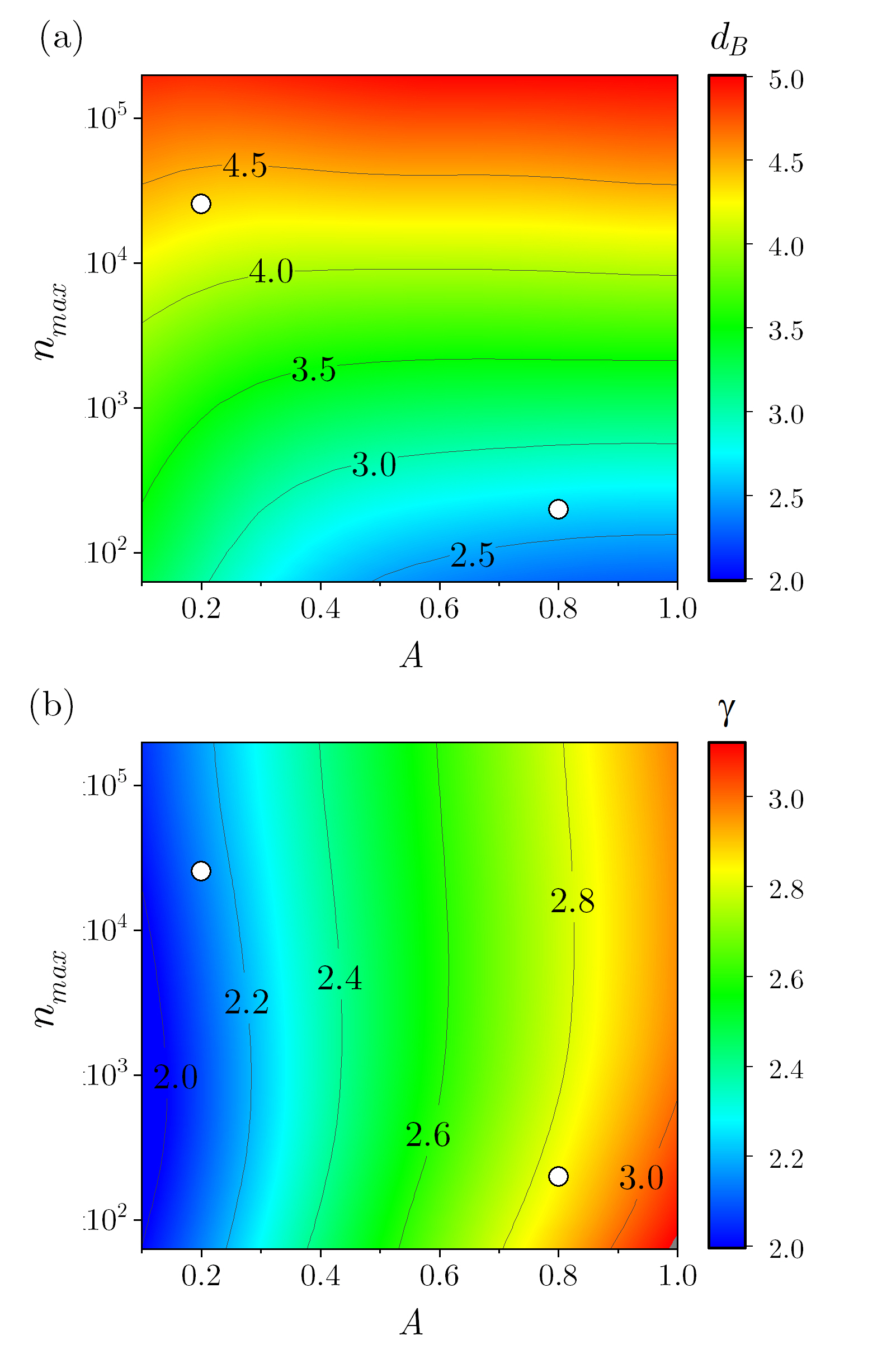}
	\caption{Heat maps illustrating the dependence of (a) the fractal dimension $d_B$ and (b) the characteristic exponent $\gamma$ of the degree distribution on the model parameters $A$ and $n_{max}$ for $a=1$ (see Fig.~S2 in SM for heat maps of standard deviation of $d_B$ and $\gamma$ and Fig.~S3 in SM for analogous heat maps for $a=2$). White circles indicate the parameter values used to generate the networks analyzed in Figs.~\ref{fig4f} and~\ref{fig5f}, where the data for each of these networks are marked with different colors. Each heat map shows results averaged over 20 network realizations with $N=10^6$, $a=1$, and $\tau=2$.}
	\label{fig3f}
\end{figure}

\begin{figure}[t]
	\centering
	\includegraphics[width=0.95\columnwidth]{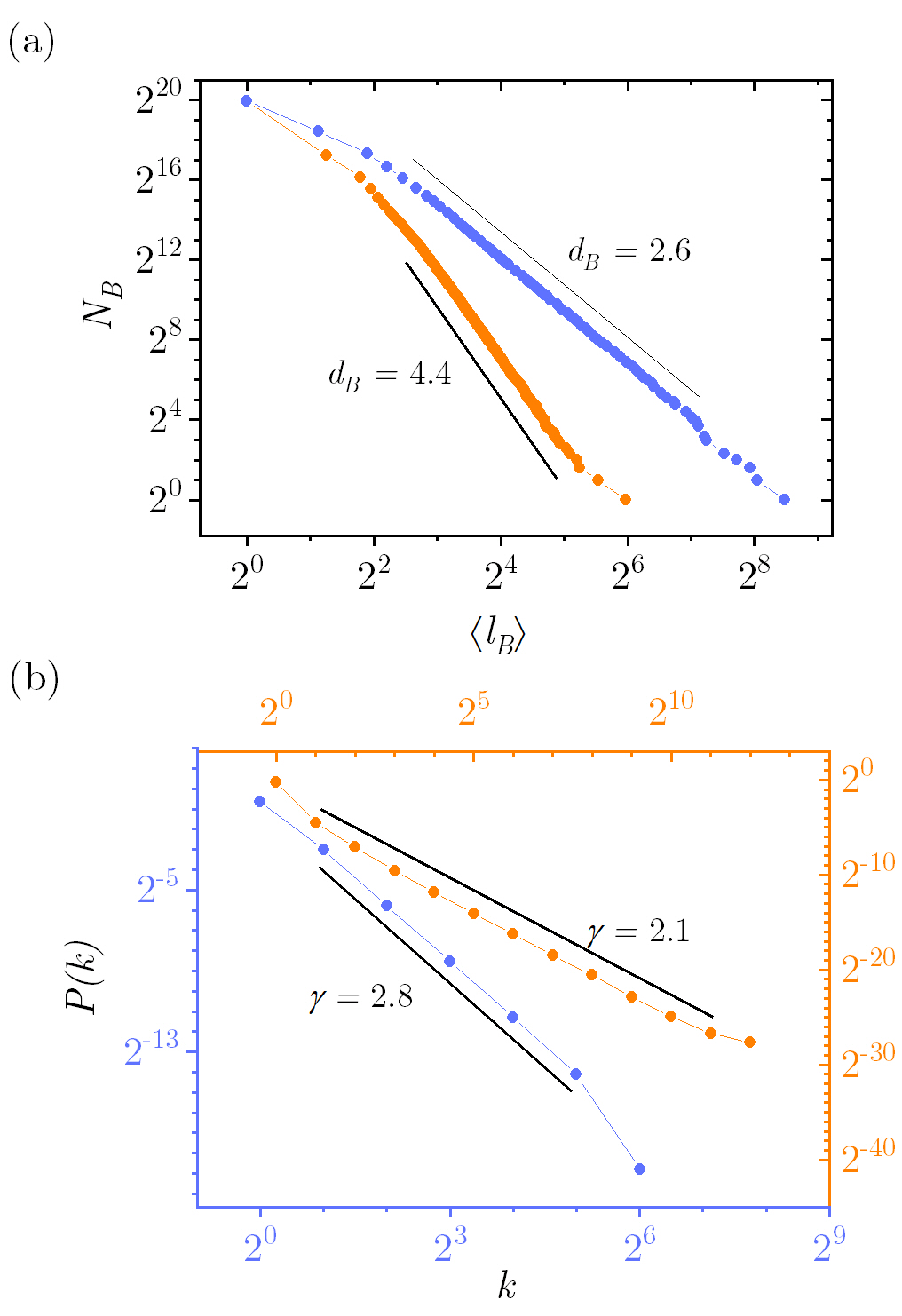}
	\caption{Examples of macroscopic characteristics of the considered model of fractal complex networks: (a) the number of boxes as a function of average box diameter, $N_B(\langle l_B \rangle)$, and (b) the node degree distribution, $P(k)$. The model parameters correspond to the white circles in Fig.~\ref{fig3f}.}
	\label{fig4f}
\end{figure}

\begin{figure}[t]
	\includegraphics[width=0.85\columnwidth]{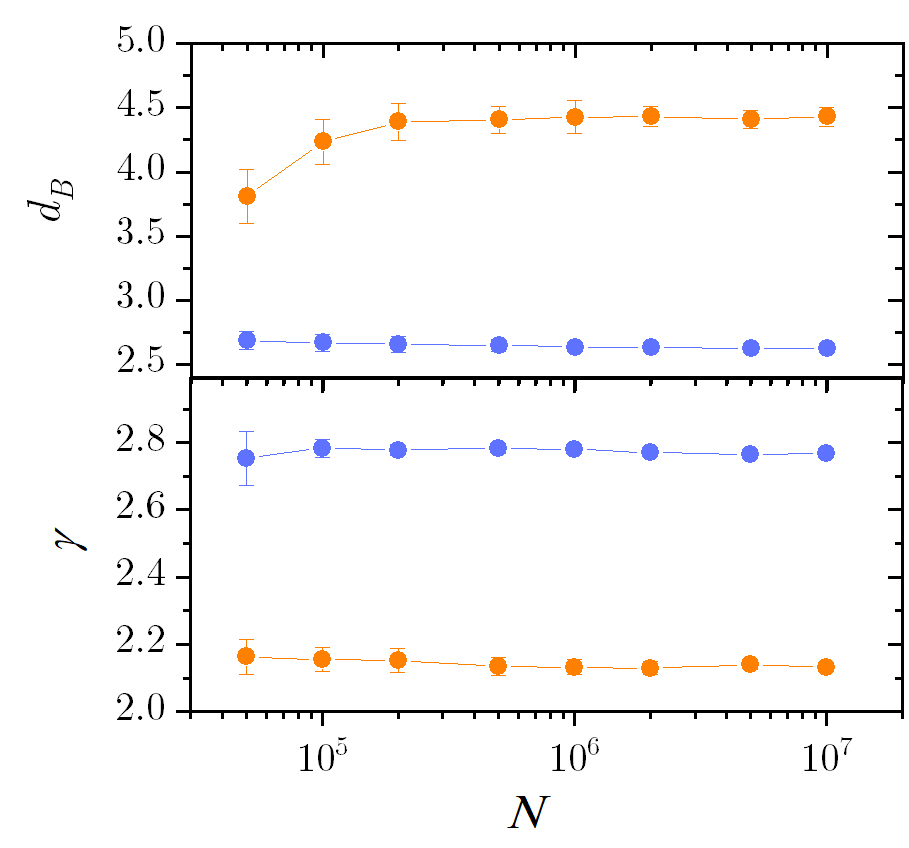}
	\caption{Dependence of the macroscopic scaling exponents $d_B$ and $\gamma$ on the network size $N$. The parameter values correspond to the white circles shown in Fig.~\ref{fig3f}. The meaning of the colors assigned to different data series is the same as in Fig.~\ref{fig4f}. Each data point represents the average over 20 independent network realizations. }
	\label{fig5f}
\end{figure}

\section{Results}\label{SecResults}

\subsection{Scale-free and self-similar properties\\of the model}\label{SecResultsA}

As shown in Figs.~\ref{fig3f}–\ref{fig5f}, the proposed model of fractal complex networks exhibits scale-free and self-similar properties over a broad range of model parameters. Below, we discuss how this parametric flexibility can be used to tune the macroscopic scaling exponents that govern fractal properties in complex networks.

In particular, Fig.~\ref{fig3f} presents two heat maps that illustrate how the fractal dimension $d_B$ and the characteristic exponent of the node degree distribution $\gamma$ depend on $A$ and $n_{max}$. The results, obtained for $N=10^6$, $a=1$, and $\tau=2$, demonstrate that the model can reproduce a wide range of scaling exponent values. Specifically, $d_B$ varies continuously from 2 to 5, while $\gamma$ ranges from 2 to 3—closely matching the variability observed in real-world fractal networks. These wide ranges are especially important, as traditional deterministic models such as SHM and $(u,v)$-flowers offer much more limited flexibility in this respect (see Tab.~1 in Ref.~\cite{Fronczak_2024}). Moreover, in those models, the scaling exponents are inherently interdependent, while in our case they can be tuned almost independently, as we demonstrate later in this section.

\begin{figure*}[th!]
	\includegraphics[width=0.99\textwidth]{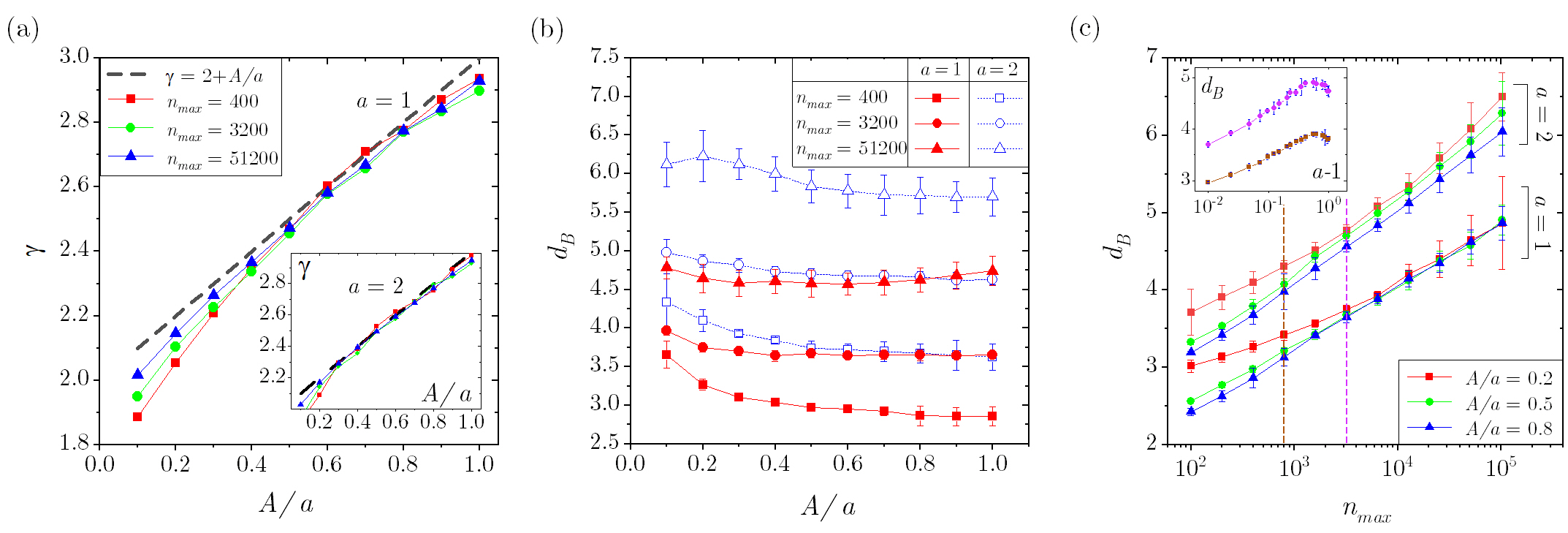}
	\caption{Overview of numerical experiments conducted to empirically determine how the characteristic exponent of the degree distribution, $\gamma$, and the box-counting dimension, $d_B$, depend on the model parameters $A$, $a$, and $n_{max}$ given $\tau=2$ (cf. the discussion of Eqs.~(\ref{gamma}) and~(\ref{dB}) in the main text).} \label{fig6f}
\end{figure*}

Before proceeding with a formal discussion of the model's properties, it is worth mentioning that generating such maps for networks of size $N=10^6$ has been made possible by a recently developed box-covering algorithm \cite{Fronczak_2025} that enables highly efficient computation of fractal dimensions even for very large networks. In contrast to standard box-covering methods \cite{Song_2007, Molontay_2021, Wen_2021}, this algorithm calculates the fractal dimension by covering networks with boxes of variable sizes. In this approach, hubs (i.e.,~nodes whose degrees exceed a given threshold) are selected as initial seeds for each box, and all remaining nodes are then assigned to their nearest hub, regardless of their exact distance. By covering the network with boxes whose number, $N_B$, is controlled via the node degree threshold, one can generate different box configurations. Calculating an average box size, $\langle l_B \rangle$, for each configuration then allows the estimation of the coverage function $N_B(\langle l_B \rangle)$, cf.~Eq.~(\ref{NblB}). As demonstrated in \cite{Fronczak_2025}, this method offers enhanced flexibility and accuracy in the analysis of fractal networks.

\begin{figure*}[th!]
	\includegraphics[width=0.99\textwidth]{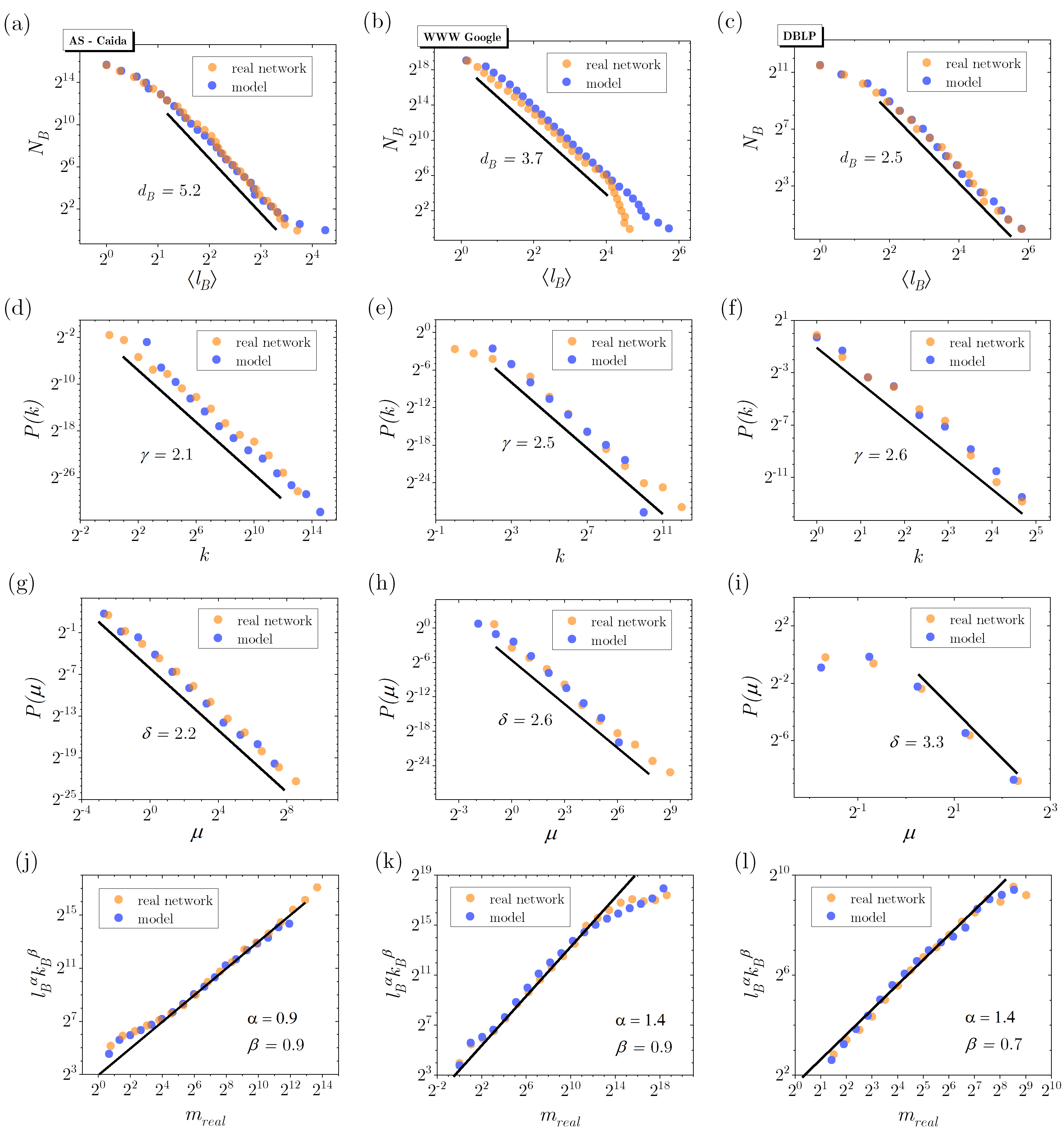}
	\caption{A comparison of the characteristics of three real-world networks (orange points) with those of networks generated using the proposed model (blue points). The graphs placed in the same column refer to the same network (i.e. Internet, WWW and DBLP, respectively). In particular, the following graphs show: (a,b,c) -- a log-log plot of $N_B$ versus $\langle l_B \rangle$ revealing the fractal nature of the network according to Eq.~(\ref{NblB}); (d,e,f) -- scale-free property of the node degree distribution $P(k)$, cf. Eq.~(\ref{Pk}); (g,h,i) -- scale-free property of the distribution of normalized box masses $P(\mu)$, cf. Eq.~(\ref{Pm}); (j,k,l) -- observed box masses versus their predicted values according to Eq.~(\ref{mlBkB}) (cf. Fig.~S4 for technical details).}
	\label{fig7f}
\end{figure*}

In Fig.~\ref{fig4f}, we present examples of two analyzed characteristics—namely the coverage function $N_B(\langle l_B\rangle)$ and the node degree distribution $P(k)$ for two selected sets of model parameters marked in Fig.~\ref{fig3f} with white circles. Complementarily, Fig.~\ref{fig5f} illustrates how the exponents $d_B$ and $\gamma$, which characterize networks generated with these parameter values, vary with the network size $N$. 

As shown in Fig.~\ref{fig5f}, the scaling exponent $\gamma$ is independent of the network size. In contrast, the fractal dimension $d_B$ initially increases with $N$, but finally stabilizes and remains constant for networks larger than $10^6$ nodes. This observation supports the assumption that the variability patterns of the scaling exponents $d_B$ and $\gamma$, as shown in Fig.~\ref{fig3f}, are representative of the considered model, regardless of $N$. Consequently, it is justified to define numerically inspired scaling functions $d_B(A,a,n_{max})$ and $\gamma(A,a,n_{max})$ for this model.

To formulate the aforementioned scaling functions, we conducted a series of numerical simulations, which are summarized in Fig.~\ref{fig6f}. In each case, we fixed the value of one of the model parameters (e.g., $n_{max}$) and examined how the scaling exponents, $\gamma$ and $d_B$, depend on the others (e.g., on the ratio $A/a$). This systematic approach, informed by the patterns observed in the heat maps, allowed us to identify the following relationships: 
\begin{equation}\label{gamma}
	\gamma(A,a)\simeq2+\frac{A}{a},
\end{equation}
and
\begin{equation}\label{dB}
	d_B(a,n_{max})\simeq\log(n_{max})+const(a).
\end{equation}

Eqs.~(\ref{gamma}) and~(\ref{dB}), which directly refer to Fig.~\ref{fig6f}, reveal a key feature of the model: For a fixed value of $a$, the parameters $A$ and $n_{max}$ independently control the scaling exponents $\gamma$ and $d_B$, respectively. This confirms a crucial property of the proposed framework—the ability to independently tune two fundamental characteristics of fractal complex networks: scale-freeness and fractality.

In particular, it is worth emphasizing that the expression for $\gamma$ given in Eq.~(\ref{gamma}) is mathematically equivalent to Eq.~(12) in Ref.~\cite{Dorogovtsev_2000}. This correspondence provides an evidence that the scale-free structure of our model is not only conceptually, but also quantitatively inherited from random networks evolving according to the g-PAR mechanism. 

In turn, Eq.~(\ref{dB}) suggests that the box dimension of the studied fractal networks depends primarily on $n_{max}$, i.e. the size of the largest non-fractal modules embedded in their structure, but also—through the parameter $a$—on the network density, which is, however, a rather non-trivial relationship that is difficult to capture in quantitative terms (see inset in Fig.~\ref{fig6f}(c), which shows the dependence of the box dimension $d_B$ on $a$ for two different values of $n_{max}$, which are marked on the main panel of the figure as dashed lines). 

Nonetheless, applying Eqs.~(\ref{gamma}) and~(\ref{dB}) to construct synthetic models of real‐world fractal networks—by inferring the parameters \(A\) and \(n_{max}\) from empirical values of \(\gamma\), \(d_B\), and \(\langle k\rangle=2a\)—requires a caution. For example, as illustrated in Fig.~\ref{fig6f}(a), in the regime \(\gamma\gtrsim2\), Eq.~(\ref{gamma}) may substantially overestimate \(\gamma\). In general, deviations from the empirical relations—Eqs.~(\ref{gamma}) and~(\ref{dB})—mainly occur for small values of $A⁄a$ and $n_{max}$, which is most likely inherited from the underlying g-PAR growth mechanism, which reaches its stationary state (characterized by a well-defined exponent $\gamma$) only in the thermodynamic limit \cite{Dorogovtsev_2000}. Such a limit is inaccessible when an upper bound is imposed on the subnetwork size, as is the case with finite $n_{max}$. Therefore, these relations are best treated as guidelines, to be refined through targeted trial‐and‐error exploration of the parameter space. 

\subsection{Model-based reconstruction of real-world networks}\label{SecResultsB}

In this section, building on the preceding results, we outline a method for selecting model parameters such that the resulting synthetic network reproduces the scaling exponents $\gamma$ and $d_B$ observed in a given real-world fractal complex network. 

The procedure involves the following steps. First, based on empirical data, one determines the network size $N$ and the parameter $a=\langle k\rangle/2$. Then, by substituting $a$ and the target value of $\gamma$ into Eq.~(\ref{gamma}), the corresponding value of the model parameter $A$ is obtained.

The second step is to determine the appropriate value of the parameter $n_{max}$, which governs the fractal properties of the network. This step is carried out in two consecutive stages. First, the additive constant in Eq.~(\ref{dB}) must be estimated. To this end, a trial numerical simulation can be performed using an arbitrary value of $n_{max}$, while keeping the remaining parameters ($N$, $a$, $A$, and $\tau$) fixed. The resulting value of $d_B$ is then used in Eq.~(\ref{dB}) to evaluate the constant. Once this constant has been established, the value of $n_{max}$ needed to reproduce the empirical fractal dimension can be calculated from Eq.~(\ref{dB}) using the observed value of $d_B$.

Also note that to accurately reproduce the characteristics of real networks, it is also necessary to adjust the $\tau$~parameter. In some networks, power-law scaling only emerges at intermediate values of $\langle l_B\rangle$, while at smaller values, the behavior is more characteristic of an exponential relationship. Adjusting parameter $\tau$ allows for a more accurate reproduction of the scaling function $N_B(\langle l_B\rangle)$—particularly in the region of small box diameters $\langle l_B\rangle$.

Fig.~\ref{fig7f} compares the macroscopic characteristics of selected real-world fractal complex networks (see Sec.~\ref{SecDatasets} for details) with those reproduced by their synthetic counterparts generated using the model proposed in this study. The parameters of both real and synthetic networks are listed in Table~\ref{tablenetworks}. 

Notably, the proposed model accurately reproduces not only the power-law behavior of the box-counting function $N_B(\langle l_B\rangle)$ (see Fig.~\ref{fig7f}(a-c)) and the scale-free nature of the degree distribution $P(k)$ (see Fig.~\ref{fig7f}(d-f)), but also the distribution of box masses $P(\mu)$ (see Fig.~\ref{fig7f}(g-i)), where $\mu = m/\langle m\rangle$ and $\langle m\rangle\simeq N/N_B(\langle l_B\rangle)=\langle l_B\rangle^{d_B}$, cf.~Eqs.~(\ref{NblB}) and~(\ref{Pm}). This, in turn, allows us to determine the characteristic exponent $\delta$ which, according to the recently developed scaling theory of fractal complex networks, together with $d_B$ and $\gamma$, forms a set of independent macroscopic scaling exponents characterizing fractal networks. 

As mentioned in Sec.~\ref{SecIntroA}, these macroscopic exponents enable the calculation of various microscopic scaling exponents, e.g.~$\alpha$ and $\beta$, which characterize the scale-invariant internal structure of boxes in fractal networks, as described by Eqs.~(\ref{ab}). The obtained values of $\alpha$ and $\beta$ can then be used to verify whether the masses of self-similar boxes—identified by the algorithm for determining the box-counting dimension—are consistent with the theoretical predictions of the scaling theory, which relates these masses to the diameters $l_B$ of the boxes and the degrees of local hubs $k_B$ within them., cf.~Eq.~(\ref{mlBkB}). This consistency test, which yields remarkably good agreement for both the real-world networks under study and their synthetic counterparts (see Fig.~\ref{fig7f}(j-l)), demonstrates that the proposed model not only reproduces the macroscopic features of real fractal networks but also captures the details of their microscopic structure.

\section{Discussion and concluding remarks}\label{SecSummary}

In this work, we introduced and numerically analyzed a new model of fractal complex networks that combines features of evolving networks and reverse renormalization. Unlike previous models—mostly deterministic in nature—the proposed approach enables flexible generation of synthetic networks that reproduce both the macroscopic and microscopic characteristics of real-world fractal systems.

One of the main advantages of the model lies in its versatility. It provides a framework in which the fundamental scaling exponents—namely, the box-counting dimension $d_B$ and the degree distribution exponent $\gamma$—can be independently tuned through interpretable parameters. This decoupling represents a notable improvement over traditional models, where these exponents tend to be tightly constrained and interdependent.

The model is not without limitations. Its most significant drawback is the lack of analytical tractability, which has not yet been resolved. This challenge is closely related to the relatively large number of parameters (five in total, including network size and average node degree). Nonetheless, these limitations appear to be outweighed by the model’s strengths—most importantly, the fact that the influence of its parameters on the resulting network structure is both interpretable and systematic. Numerical experiments confirm that the key macroscopic scaling exponents respond to parameter changes in a predictable way, which allows for accurate, targeted reconstruction of real fractal networks.

Another limitation of the model in its current form is the lack of independent control over the exponent \(\delta\), which governs the scale‐free behavior of the box‐mass distribution. Although this issue was deliberately set aside in our analysis of the model’s core properties, it is noteworthy that the proposed framework nonetheless reproduces the empirical box‐mass distributions in every real‐world network examined. On one hand, this remarkable agreement could suggest that our model captures the essential mechanisms underlying real fractal complex networks with exceptional accuracy. On the other hand, prudence dictates that these matching distributions may simply reflect as‐yet‐undiscovered scaling relations in fractal networks rather than direct evidence of our model’s completeness.

In our recent paper on the scaling theory of fractal complex networks \cite{Fronczak_2024}, we formulated novel relations linking macroscopic scaling exponents characterizing structural properties ($d_B$, $\gamma$, and $\delta$) to microscopic ones (including $\alpha$ and $\beta$). That theory, however, does not address network density or edge‐scaling. Edge‐scaling analysis is important for several reasons \cite{Gallos_2007}. From a theoretical standpoint, understanding how the number of edges grows within renormalization boxes could reveal additional macroscopic exponents that complete the scaling framework of fractal networks. From a simulation‐oriented perspective, refined edge‐scaling laws are crucial for studying dynamical processes—such as information diffusion or epidemic spreading—whose behavior depends sensitively on local connectivity patterns. To capture and test these effects, models must support flexible edge‐scaling rules and thereby provide a reliable platform for exploring the interplay between topology and dynamics. We believe that the fractal network model examined in this work meets these requirements.

Finally, we turn to the recently suggested links between fractality and modularity. Although our nested‐subnetwork architecture already produces moderate clustering at small scales, further enhancements remain possible. For example, replacing the standard g-PAR mechanism with growth rules explicitly designed to encourage triangle formation—such as those in the Holme–Kim \cite{Holme_2002} or Klemm–Eguíluz \cite{Klemm_2002} models—could significantly increase the clustering coefficient. This extension would be particularly valuable for probing the relationship between fractality and hierarchical community structure \cite{Blondel_2008}, an interconnection highlighted in our recent study \cite{Samsel_2023}. Notably, existing benchmarks for community detection (e.g., the LFR model \cite{LFRmodel} or the exponential random graph (ERG) benchmark~\cite{Fronczak_2013, Kowalczyk_2017}) lack genuine fractal organization, and classical fractal models—with their locally tree-like topologies—are ill-suited for evaluating modularity algorithms. In contrast, our proposed model, even in its current form, satisfies the criteria for both fractal and community-based benchmarking, offering a unique testbed for assessing community detection methods on truly fractal networks.

\section*{Info about supplementary materials}

In the supplementary materials, we have included additional analyses conducted as part of our study of the proposed model. Specifically:
\begin{itemize}
	\item we discussed the mapping function $f(k_i)$ (\ref{fk}) and justified the choice of the parameter $\tau = 2$;
	\item we explained the procedure for constructing networks for non-integer values of the parameter $a$;
	\item we presented additional heatmaps for $a = 2$ as well as maps illustrating the uncertainty associated with the estimated network parameters, $d_B$ and $\gamma$;
	\item we provided a detailed illustration of the methodology used to extract the relevant scaling exponents from the networks generated by the model.
\end{itemize}

In \cite{Mendeley}, we have also provided Python code for generating networks according to the presented model, as well as tools for analyzing the obtained network, including the mentioned box-covering algorithm. The code was written with a focus on efficiency at the expense of readability. This allows for the generation and analysis of large networks within a reasonable time (e.g., generating a network with \(N = 10^6\) on a standard laptop takes approximately 15 seconds, while the box-covering process adds another 60 seconds).
 
\section*{Acknowledgments}
Research was funded by Warsaw University of Technology within the Excellence Initiative: Research University (IDUB) programme.


\begin{thebibliography}{99}
	
	
\bibitem{Song_2005}
C.~Song, S.~Havlin, H.A.~Makse, \textit{Self-similarity of complex networks}, Nature {\bf 433}, 392 (2005).

\bibitem{Song_2006}
C.~Song, S.~Havlin, H.A.~Makse, \textit{Origins of fractality in the growth of complex networks}, Nat. Phys. {\bf 2}, 275 (2006).

\bibitem{Fronczak_2024}
A. Fronczak, P. Fronczak, M.J. Samsel, et al., \textit{Scaling theory of fractal complex networks}, Sci. Rep. {\bf 14}, 9079 (2024).

\bibitem{Gallos_2012}
L.K. Gallos, H.A. Makse, M. Sigman,  \textit{A small world of weak ties provides optimal global integration of self-similar modules in functional brain networks}, Proc. Natl. Acad. Sci. U.S.A. {\bf 109}, 2825 (2012). 

\bibitem{Gallos_2013}
L.K. Gallos, F.Q. Potiguar, J.S. Andrade Jr, H.A. Makse, \textit{IMDB network revisited: Unveiling fractal and modular properties from a typical small-world network}, PLOS ONE \textbf{8}, e66443 (2013).

\bibitem{Fronczak_2022} 
A. Fronczak, M.J. Mrowinski, P. Fronczak, \textit{Scientific success from the perspective of the strength of weak ties}, Sci. Rep. {\bf 12}, 5074 (2022). 

\bibitem{Gallos_2007}
L.K. Gallos, C. Song, S. Havlin, H.A. Makse, \textit{Scaling theory of transport in complex biological networks}, Proc. Natl. Acad. Sci. U.S.A. \textbf{104}, 7746 (2007).

\bibitem{Chen_2019}
Y. Wu, Z. Chen, K. Yao, X. Zhao , Y. Chen, \textit{On the correlation between fractal dimension and robustness of complex networks}, Fractals \textbf{27},  1950067 (2019).

\bibitem{Goh_2006}
K.-I. Goh, G. Salvi, B. Kahng, D. Kim, \textit{Skeleton and fractal scaling in complex networks}, Phys. Rev. Lett. \textbf{96}, 018701 (2006).

\bibitem{Deppman_2021}
A. Deppman, E.O. Andrade-Ii, \textit{Emergency of Tsallis statistics in fractal networks}, PLoS One \textbf{16}, e0257855 (2021).

\bibitem{Rozenfeld_2010}
H.D. Rozenfeld, C. Song, H.A. Makse, \textit{Small-world to fractal transition in complex networks: A renormalization group approach}, Phys. Rev. Lett. {\bf 104}, 025701 (2010).

\bibitem{SW1} M.E.J. Newman, S.H. Strogatz, D.J. Watts, \textit{Random graphs with arbitrary degree distributions and their applications}, Phys. Rev. E \textbf{64}, 026118 (2001).

\bibitem{SW2} R. Cohen, S. Havlin, \textit{Scale-free networks are ultrasmall}, Phys. Rev. Lett. \textbf{90}, 058701 (2003). 

\bibitem{SW3} A. Fronczak, P. Fronczak, J.A. Ho\l{}yst, \textit{Average path length in random networks}, Phys. Rev. E \textbf{70}, 056110 (2004).

\bibitem{road_2021}
H. Zhang, P. Gao, T. Lan, C. Liu, \textit{Exploring the structural fractality of urban road networks by different representations}, Prof. Geogr. \textbf{73}, 348 (2021).

\bibitem{BAnet} 
A.-L. Barab\'asi, R. Albert, \textit{Emergence of scaling in random networks}, Science \textbf{286}, 509 (1999).

\bibitem{Radicchi_2008}
F. Radicchi, J.J. Ramasco, A. Barrat, S. Fortunato, \textit{Complex networks renormalization: flows and fixed points}, Phys. Rev. Lett. \textbf{101}, 148701 (2008).

\bibitem{Rosenberg_2020_book}
E.~Rosenberg, {\em Fractal Dimensions of Networks}, Springer (2020).

\bibitem{Yook_2005} S.-H. Yook, F. Radicchi, H. Meyer-Ortmanns, \textit{Self-similar scale-free networks and disassortativity}, Phys. Rev. E \textbf{72}, 045105(R)(2005).

\bibitem{Zeng_2017_sierpinski}
C. Zeng, M. Zhou, \textit{Small-world and scale-free properties of fractal networks modeled on n-dimensional Sierpinski pyramid}, Fractals \textbf{25}, 1750057 (2017).

\bibitem{Zeng_2021_sturmian}
C. Zeng, Y. Xue, Y. Huang, \textit{Fractal networks with Sturmian structure}, Phys. A: Stat. Mech. Appl. \textbf{574}, 125977 (2021).

\bibitem{Le_2015_sierpinski}
A. Le, F. Gao, L. Xi, S. Yin, \textit{Complex networks modeled on the Sierpinski gasket}, Phys. A: Stat. Mech. Appl. \textbf{436}, 646 (2015).

\bibitem{Huang_2023_pentagon}
L. Huang, Y. Zheng, \textit{Asymptotic formula on APL of fractal evolving networks generated by Durer Pentagon}, Chaos Solitons Fractals \textbf{167}, 113042 (2023).

\bibitem{Chelminiak_2013}
P. Che\l miniak, \textit{Emergence of fractal scale-free networks from stochastic evolution on the cayley tree}, Phys. Lett.~A \textbf{377}, 2846 (2013).

\bibitem{Ikeda_2020}
N. Ikeda, \textit{Fractal networks induced by movements of random walkers on a tree graph}, Phys.~A: Stat. Mech. Appl. \textbf{537}, 122743 (2020).

\bibitem{Rozenfeld_2007}
H.D. Rozenfeld, S. Havlin, D. Ben-Avraham, \textit{Fractal and transfractal recursive scale-free nets}, New. J. Phys. {\bf 9}, 175 (2007).

\bibitem{Zakar_2022}
E. Zakar-Polyak, M. Nagy, R. Molontay, \textit{Investigating the origins of fractality based on two novel fractal network models}, in Complex Networks XIII (D. Pacheco, A. S. Teixeira, H. Barbosa, R. Menezes, and G. Mangioni, eds.), (Cham), pp. 43–54, Springer International Publishing, 2022.

\bibitem{Yakubo_2022}
K. Yakubo, Y. Fujiki, \textit{A general model of hierarchical fractal scale-free networks}, PLOS ONE \textbf{17}, e0264589 (2022).

\bibitem{Dorogovtsev_2000}
S.N. Dorogovtsev, J.F.F. Mendes, A.N. Samukhin, \textit{Structure of growing networks with preferential linking}, Phys. Rev. Lett. {\bf 85}, 4633 (2000).

\bibitem{caida_as}
Center for Applied Internet Data Analysis, https://www.caida.org/catalog/datasets/as-relationships/, accessed: 2024-10-04.

\bibitem{routeviews}
University of Oregon RouteViews Project, https://www.routeviews.org/, accessed: 2024-10-04.

\bibitem{stanford_caida_repo}
Stanford Large Network Dataset Collection, https://snap.stanford.edu/data/as-Caida.html, accessed: 2024-10-04.

\bibitem{Leskovec_2009} 
J. Leskovec, K.J. Lang, A. Dasgupta, M.W. Mahoney, \textit{Community structure in large networks: Natural cluster sizes and the absence of large well-defined clusters}, Internet Math. {\bf 6}, 29 (2009).

\bibitem{Rossi_2015_repo}
R.A. Rossi, N.K. Ahmed, \textit{The network data repository with interactive graph analytics and visualization}, Proc. AAAI Conf. Artificial. Intell.  {\bf 29}(1) (2015), https://networkrepository.com.

\bibitem{Tang_2012}
J. Tang, A.C.M. Fong, B. Wang, J. Zhang,\textit{ A unified probabilistic framework for name disambiguation in digital library}, IEEE Trans. Knowl. Data Eng. {\bf 24}, 975 (2012).

\bibitem{DBLP_repo}
DBLP Citation Network Dataset, https://www.aminer.org/citation, accessed: 2022-08-30.
	
\bibitem{Fronczak_2025}
M. \L{}epek, K. Makulski, A. Fronczak, P. Fronczak, \textit{Beyond traditional box-covering: Determining the fractal dimension of complex networks using a fixed number of boxes of flexible diameter}, arXiv:2501.16030.

\bibitem{Song_2007}
C. Song, L.K. Gallos, S. Havlin, H.A. Makse, \textit{How to calculate the fractal dimension of a complex network: the box covering algorithm}, J. Stat. Mech.: Theory Exp. P03006 (2007).

\bibitem{Molontay_2021}
P.T. Kov\'{a}cs, M. Nagy, R. Molontay, \textit{Comparative analysis of box-covering algorithms for fractal networks}, Appl. Netw. Sci. {\bf 6}, 73 (2021).

\bibitem{Wen_2021}
T.~Wen, K.H.~Cheong, \textit{The fractal dimension of complex networks: a review}, Inf. Fusion {\bf 73}, 87 (2021).

\bibitem{Holme_2002}
P. Holme, B.J. Kim, \textit{Growing scale-free networks with tunable clustering}, Phys. Rev. E \textbf{65}, 026107 (2002).

\bibitem{Klemm_2002}
K. Klemm, V.M. Equ\'iluz, \textit{Highly clustered scale-free networks}, Phys. Rev. E \textbf{65}, 036123 (2002).

\bibitem{Blondel_2008} 
V.D. Blondel, J.-L. Guillaume, R. Lambiotte, E. Lefebvre, \textit{Fast unfolding of communities in large networks}, J. Stat. Mech.: Theory Exp. P10008 (2008).

\bibitem{Samsel_2023}
M.J. Samsel, K. Makulski, M. \L{}epek, A. Fronczak, P. Fronczak, \textit{Towards fractal origins of the community structure in complex networks: a model-based approach}, arXiv:2309.11126.

\bibitem{LFRmodel}
A. Lancichinetti, S. Fortunato, F. Radicchi, \textit{Benchmark graphs for testing community detection algorithms}, Phys. Rev. E \textbf{78}, 046110 (2008).

\bibitem{Fronczak_2013} P. Fronczak, A. Fronczak, M. Bujok, \textit{Exponential random graph models for networks with community structure}, Phys. Rev. E \textbf{88}, 032810 (2013).

\bibitem{Kowalczyk_2017} M. Kowalczyk, P. Fronczak, A. Fronczak, \textit{A simple and efficient algorithm for modeling modular complex networks}, Phys.~A: Stat. Mech. Appl. \textbf{482}, 218 (2017).

\bibitem{Mendeley} P. Fronczak, Mendeley Data, V1, doi: 10.17632/9t8gkfj428.1 (2025).

\end{thebibliography}
\end{document}